# Numerical Characterization of *In Vivo* Wireless Communication Channels

A. Fatih Demir[1], Qammer H. Abbasi[2], Z. Esad Ankarali[1], Erchin Serpedin[2], Huseyin Arslan[1,3]

[1]Department of Electrical Engineering, University of South Florida, Tampa FL, USA
[2]Department of Electrical and Computer Engineering, Texas A&M University, USA/Qatar
[3] Department of Electrical and Electronics Engineering, Istanbul Medipol University, Istanbul, Turkey
Email: {afdemir, zekeriyya}@mail.usf.edu, qammer.abbasi@tamu.edu, serpedin@ece.tamu.edu, arslan@usf.edu

*Abstract*—In this paper, we numerically investigated the *in vivo* wireless communication channel for the human male torso at 915 MHz. The results show that *in vivo* channel is different from the classical communication channel, and location dependency is very critical for link budget calculations. A statistical path loss model based on the angle, depth and body region is introduced for near and far field regions. Furthermore, multipath characteristics are investigated using a power delay profile as well.

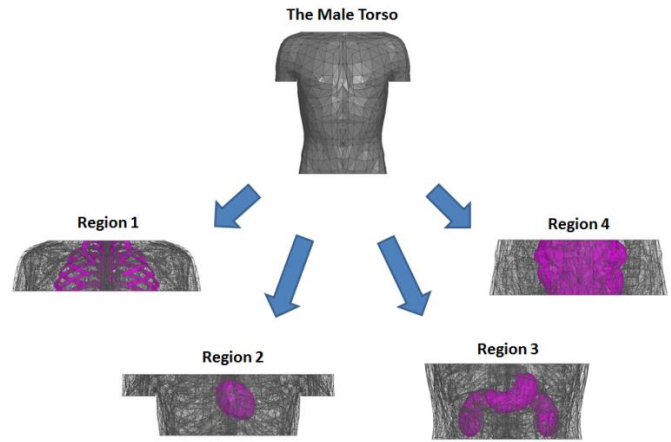

Fig. 1: Regions 1 to 4 represent the body areas of shoulders to heart (including first six rib bones), heart, stomach - kidneys and intestine respectively.

## I. INTRODUCTION

Technological advances in biomedical engineering have increased the quality of life and life expectancy of people. Early diagnosis of diseases is crucial, but the cost of healthcare is not affordable for everyone. Wireless *in vivo* sensors and actuators offer cost efficient solution for this problem providing a reliable continuous monitoring system of patient's vital signs. Meanwhile, deployment of these devices reduces hospital visits and invasiveness of surgeries [1].

Accurate channel models are important to optimize physical layer communication system in terms of power and spectral efficiency. Although on body wireless communication channel models are well studied [2], there are relatively fewer studies in the literature for the *in vivo* wireless communication channels. The *in vivo* channel characteristics are different from the classical communication channel characteristics due to electromagnetic wave propagation through various tissues and organs, which have different electrical properties. The liquid nature of human body structure causes RF attenuation, whereas, the skeletal structure causes wave diffraction and refraction with respect to corresponding frequency [3].

There is a tremendous ongoing research to characterize *in vivo* wireless communication channel in recent years [3]-[8]. Sani *et al.* [4] demonstrated that radiation characteristics of wireless implants are subject specific and strongly related to the location of the sensor. Ketterl *at al.* [5] showed that even slight changes in the position of the external receive antenna at a fixed distance from the internal antenna cause significant variations in received signal strength. Therefore channel models without considering the relevant body region may not be reliable. In [6], it was found that there is no direct co-relation between near and far fields due to the presence of reactive fields in the near region of ingested sources. Moreover, Alomainy *et al.* [7] revealed that electric field attenuation is more dependent on distance rather than frequency because of near-field effect inside the empty stomach. These results show that both near and far fields need to be investigated to obtain sufficient information about *in vivo* channel characteristics.

This paper presents a detailed numerical characterization (by performing approximately 1500 simulations) of *in vivo* wireless communication channel for human male torso by considering location dependency (angle, depth and body region) for both near and far field regions. Based on these parameters a statistical formula for the path loss is defined, and the multipath characteristics of the channel are examined through a power delay profile.

## II. SIMULATION SETUP

Analytical methods are infeasible and require extreme simplifications for radio propagation investigation inside the human body, and hence numerical methods are usually performed [3] – [8]. In this paper, we used ANSYS HFSS 15.0, which is a full-wave electromagnetic field simulator based on Finite Element Method (FEM). ANSYS also provides a detailed human body model of an adult male.

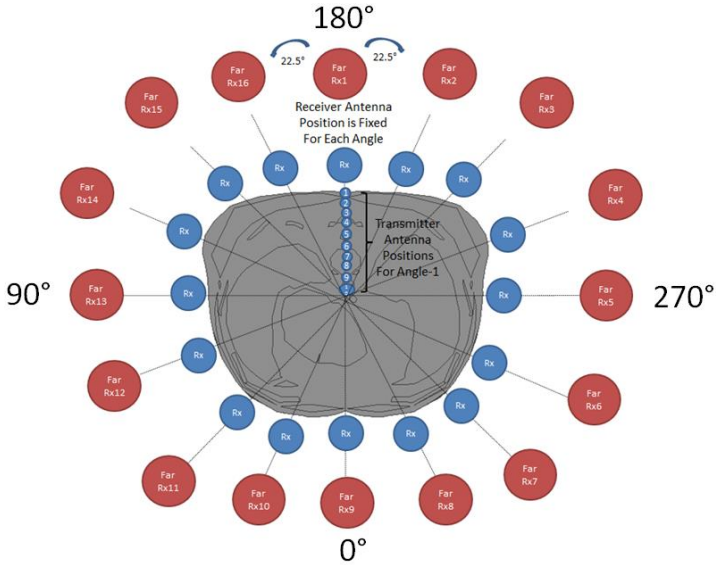

Fig. 2: 16 (angles) x 10 (depth) x 2 (near-far regions) x 4(subregions) = 1280 simulations in total for path loss.

Due to the inhomogeneous characteristics of the human body, investigation of the wireless communication channel without taking into account the relevant body area would be insufficient. Hence, torso area is divided into four subregions considering major internal organs like heart, stomach, kidneys and intestine as shown in Fig. 1. The measurements are taken for each subregion by rotating both receiver (*ex vivo*) and transmitter antennas (*in vivo*) around the body in X-Y plane with 22.5° angle increments as shown in Fig. 2. For a fixed position of receiver antenna, the transmitter antenna is placed at ten different depth values (10mm to 100mm). It is important to point out that antennas are not rotating in a circular path around the human body. In each angle, a new coordinate system is defined exactly on the body surface, and both antennas are placed with respect to this new coordinate system. Moreover, antennas are placed in the same orientation so that there is no polarization loss in the results.

Omni-directional dipole antennas at 915 MHz (ISM band) are used to take measurements in the simulations. A dipole antenna's size is proportional to the wavelength, which changes with respect to both frequency and permittivity. Thus, the *in vivo* antennas typically have geometry smaller than in the free space, which causes very low transmission efficiency. However, the characteristics of the biological tissues make small antennas be an efficient radiator [3]. In this study, operation frequency is fixed but permittivity value of the environment is variable. Therefore, average permittivity value of each body region is estimated, and then appropriate antenna size is selected for maximum power delivery. The estimated average permittivity values of Region 1 to 4 are 16, 16, 27 and 29 respectively. Near and far field scenarios are investigated by placing the receiver antenna (*ex vivo*) 5 cm and 30 cm away from the body surface, respectively. Antenna locations with high return loss (i.e. >-7dB) are discarded from the data.

## III. RADIO CHANNEL CHARACTERIZATION

We investigated the angle, depth and body region dependent characteristic of path loss, which is a measure of average signal power attenuation, for near and far field regions. The signal propagates through different organs and tissues for various antenna locations, and hence the path loss changes significantly even for the same depth value from the body surface. Fig. 3 shows results of near-field path loss for each angle, where linear averages of four region's path loss values are calculated for each depth and angle (far-field figures are not included in the paper due to limited space). It is observed that 0° has the highest path loss, whereas symmetric locations, 112.5° and 247.5° have the lowest attenuation. The number of scattering objects (random variables) increases as the *in vivo* antenna goes deeper. Therefore, the variance of each depth's path loss values among 16 angles increases as the increase in depth due to the summation of random variables as can be seen in Fig.3.

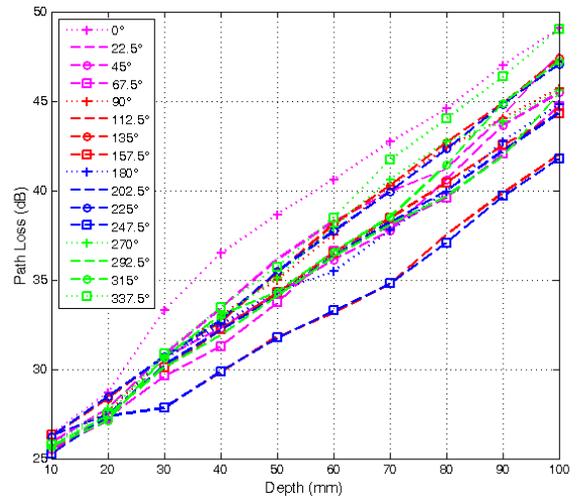

Fig. 3: Average path loss for each angle at 915 MHz.

In addition, the path loss values are fitted into a statistical formula using least squares linear regression model. In the literature, the path loss is commonly modeled with a logarithmic function [1], [4], [7], [8], however, we observed that a linear function fits better since the convergence of mean square error is smaller when the gradient descent algorithm is applied. The *in vivo* path loss is modeled as a function of depth by the following linear equation in dB:

$$PL(d) = PL_0 + m(d/d_0) + S \qquad (d \geq d_0)$$

where $d$ is the depth distance from body surface in millimeters, $d_0$ is the reference depth distance (i.e. 10mm), $PL_0$ is the intersection term in dB, $m$ is the decay rate of received power and $S$ is the random scattering parameter in dB, which is a normally distributed random variable with zero mean and variance $\sigma$. These random scatters caused by different body materials and the antenna gain in different directions [8].

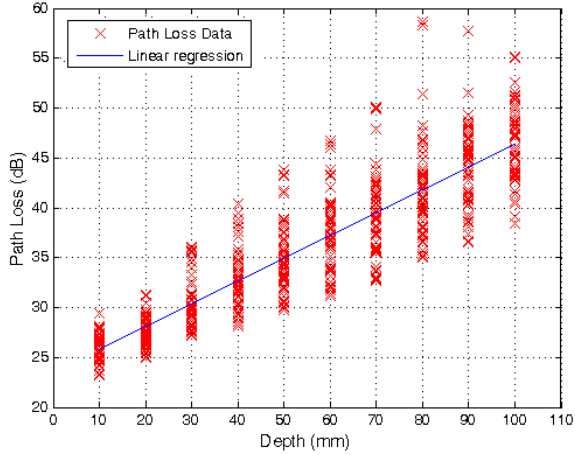

Fig. 4: Path loss versus distance for near-field setup at 915 MHz.

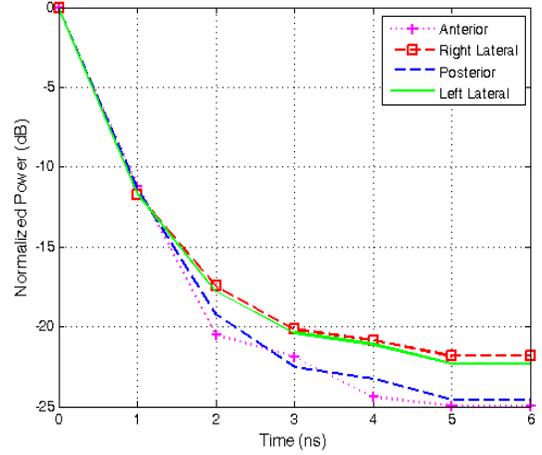

Fig. 5: Power delay profile for each anatomical direction.

The statistical parameters for path loss model are summarized in Table 1. There is a 30% difference in the decay rate of received power (*m*) between Region 2 and Region 3. The path loss at Region 1 and Region 4 show more deviation around the mean than other two regions. The angle values are combined into four sides to show angle dependency of the channel in a more convenient way. The posterior body has the least decay rate and the variation in path loss, whereas anterior body has the highest values. Additionally, the path loss decay rate is higher in near-field measurements than the far field for all regions and sides as expected. The multipath characteristics of the channel can be investigated by a power delay profile (Fig. 5), and greater dispersion is observed on the sides than the anterior or posterior body. These variations based on the angle, depth, and body region are required to be considered for link budget calculations not to harm biological tissues and to provide power efficiency.

## IV. CONCLUSION

We presented location dependent characteristics of the *in vivo* wireless communication channel using a full wave numerical RF simulator for both near and far field regions. A statistical path loss formula based on the angle, depth, and body region parameters is introduced, and the multipath characteristics are investigated for various antenna locations. These results emphasize the difference of *in vivo* wireless communication channels than the classical channels and show that the *in vivo* wireless communication systems need to be designed carefully with regards to the location of operation.

## ACKNOWLEDGEMENT

This publication was made possible by NPRP grant # NPRP 6 - 415 - 3 - 111 from the Qatar National Research Fund (a member of Qatar Foundation). The statements made herein are solely the responsibility of the authors.

TABLE 1: PARAMETERS FOR THE STATISTICAL PATH LOSS

| Parameters / Body Area | Near Field $PL_0$[dB] | Near Field m | Near Field σ | Far Field $PL_0$[dB] | Far Field m | Far Field σ |
|---|---|---|---|---|---|---|
| Region1 | 24.75 | 2.30 | 3.73 | 41.07 | 1.46 | 2.84 |
| Region2 | 22.70 | 1.96 | 2.38 | 39.37 | 1.48 | 3.04 |
| Region3 | 22.56 | 2.55 | 1.79 | 39.09 | 2.14 | 3.02 |
| Region4 | 24.23 | 2.31 | 3.47 | 41.05 | 1.82 | 3.90 |
| Anterior | 23.83 | 2.46 | 3.51 | 40.57 | 1.88 | 4.08 |
| Posterior | 23.76 | 2.21 | 1.92 | 40.53 | 1.76 | 2.34 |
| Left Lateral | 23.34 | 2.28 | 3.67 | 39.57 | 1.68 | 3.62 |
| Right Lateral | 23.22 | 2.27 | 3.51 | 39.43 | 1.69 | 3.52 |
| Overall Torso Area | 23.56 | 2.28 | 3.38 | 40.14 | 1.73 | 3.62 |

3